\documentstyle[eqsecnum,aps,epsf]{revtex} 
\newcommand{\postscript}[2] {\setlength{\epsfxsize}{#2\hsize}
\centerline{\epsfbox{#1}}}

\begin{document}
\twocolumn[\hsize\textwidth\columnwidth\hsize\csname@twocolumnfalse\endcsname

\title{Limits of validity for a semiclassical mean-field two-fluid model
for Bose-Einstein
condensation thermodynamics}

\author{Sadhan K. Adhikari\footnote{E-mail: adhikari@ift.unesp.br} and A.
Gammal\footnote{E-mail: gammal@ift.unesp.br}}
\address{Instituto de F\'{\i}sica Te\'orica, Universidade Estadual Paulista,
01.405-900 S\~ao Paulo, S\~ao Paulo, Brazil\\}

\date{\today}
\maketitle
\begin{abstract}

We reinvestigate the Bose-Einstein condensation (BEC) thermodynamics of a
weakly
interacting dilute Bose gas under the action of a trap using a 
semiclassical two-fluid mean-field model in order to find the domain of
applicability of the model. Such a  model is expected to  break down once
the condition of diluteness and weak interaction is violated.  We find
that
this breakdown  happens
for values of coupling and density near the 
present experimental scenario of BEC.
With the increase of the interaction coupling and density the
 model  may lead to
unphysical results for thermodynamic observables.

{\bf PACS Number(s): 03.75Fi, 05.30.Jp,  67.40Kh} 

\end{abstract} 
\vskip1.5pc] 

\section{Introduction}
The recent experimental observation of Bose-Einstein  condensation
(BEC) in
a weakly interacting dilute gas of $^{87}$Rb \cite{1}, $^{23}$Na \cite{2},
$^7$Li \cite{3}, and $^1$H \cite{4} employing magnetic traps at ultra-low
temperatures calls for a theoretical investigation on various aspects of
the condensate. The condensate consists of few thousands to few millions
of atoms confined by the trap potential. As the temperature is lowered
below the critical temperature $T_0$ of BEC, the condensate
starts to form and finally at 0 K all the available atoms will be
condensed. In the absence of a microscopic equation, the condensate is
usually described by the mean-field Gross-Pitaevskii (GP)  equation
\cite{5}.  One of the primary interest on the process is to study the
 thermodynamic observables of the system,
such as, the condensate fraction, internal energy, and specific heat.

There have been several comprehensive studies on temperature dependencies
of thermodynamic observables of the condensate 
using semiclassical mean-field two-fluid models \cite{6,6a,7,9,10,14}. The
physical ingredients of these models are quite similar. One such model
using the GP wave function provides
satisfactory description of the temperature dependencies of the
thermodynamic observables in  two \cite{9}, and three
\cite{7,7a} dimensions. These studies employed a iterative solution of the
system of equations involved. For the condensation of a system composed of
40000 trapped $^{87}$Rb atoms with repulsive interatomic interaction the
iterative scheme converged rapidly and provided a satisfactory account of
the condensate fraction, internal energy, and 
specific heat \cite{7}.
Similar conclusion was also reached in the study of condensation of $^7$Li
atoms \cite{7a}. In case of $^7$Li the attractive interatomic interaction
is responsible for collapse if the number of atoms is larger than
approximately 1400
\cite{7b}.

Here we reinvestigate critically the BEC of a weakly interacting dilute
gas in two and three dimensions using the two-fluid mean-field model
mentioned above in order to define the domain of its applicability.  We
have included the two-dimensional BEC in this study because of
considerable recent interest in this topic \cite{9,11,12,12a,13}.  We
employ the usual iterative solution of the nonlinear two-fluid mean-field
model and study the convergence of the iterative scheme. Although, the
convergence is rapid for a weakly interacting dilute system, with an
increase of the strength of interaction and/or density, the model breaks
down and leads to physically unacceptable results for the thermodynamic
observables.  Specifically, below the critical temperature, the model may
yield negative specific heat.

However, it is well-known that the mean-field description of the
condensate via the GP equation as used in the semiclassical models above
should hold under the condition of diluteness of a weakly interacting Bose
gas and is expected to break down once the conditions of diluteness and
weak interaction are violated 
\footnote{See, for example, page 474 of Ref. \cite{14}}. The breakdown
should happen for a large number of condensed
particles reflecting a large density as well as for a large modulus of the
scattering length denoting a strong interatomic interaction. These two
conditions correspond to a large nonlinearity of the system and may lead
to a breakdown of the mean-field two-fluid thermodynamic model
\cite{6,14}.
In addition,
the finite size of the system may necessitate corrections in the
thermodynamic model as we are away from the real thermodynamic limit 
$N \to \infty$, $V\to \infty$, where $N$  is the number of particles
and $V$ the volume of the system \footnote{These limitations of the
thermodynamic model 
are discussed in Sec. 5 of
Ref.\cite{14}}.   
Nevertheless exact numerical conditions for the breakdown of the
mean-field
thermodynamical model 
have never been
investigated. In this work by performing numerical calculations we
identify such
conditions. We find that the semiclassical model may break down under
possible present experimental conditions of BEC of a trapped Bose gas.

We present the semiclassical model in Sec. 2, numerical results in Sec.
3, and conclusions in Sec. 4.   

\section{Mean-field Model}

We consider a system of $N$ bosons  with repulsive interaction at
temperature $T$
under the influence of a trap potential. The condensate is described by
the following GP wave function in the Thomas-Fermi approximation
\cite{7,9}:  \begin{equation}\label{1}
 |\Psi(r)|^2=\frac{\mu-V_{\mbox{ext}}-2gn_1(r)}{g}
\theta(\mu-V_{\mbox{ext}}-2gn_1(r)), \end{equation} where $\theta(x)$ is
the step function, $\theta(x) =0$ if $x<0$ and 1 otherwise. Here
$V_{\mbox{ext}}(r)\equiv m\omega^2r^2/2$ is the trap potential, $g$ the
strength of the repulsive interaction between the atoms, $m$ the mass of a
single bosonic atom, $\omega$ the angular frequency, $\mu$ the chemical
potential, and $n_1(r)$ represents the distribution function of the
noncondensed bosons. As we are interested in studying the limits of
validity of the semiclassical mean-field model and not in simulating a
particular experimental situation, we consider a spherically symmetric
trap both in two and three dimensions. The noncondensed particles are
treated as non-interacting bosons in an effective potential \cite{16}
\begin{equation}\label{2a} V_{\mbox{eff}}(r )=V_{\mbox{ext}}(r) 
+2gn_1(r)+2g|\Psi(r)| ^2.\end{equation} Thermal averages are calculated
with
a standard Bose distribution of the noncondensed particles in chemical
equilibrium with the condensate governed by the same chemical potential
$\mu$. In particular the density $n_1(r)$ is given by \cite{7,9,16}
\begin{equation}\label{2} n_1(r) = \frac{1}{(2\pi \hbar)^{\cal
D}}\int\frac{d^{\cal D}p}{\exp[\{p^2/2m+V_{\mbox{eff}}(r)-\mu\} /k_BT]-1},
\end{equation} where $k_B$ is the Boltzmann constant, and ${\cal D}
(\equiv 2,3)$ is the dimension of space.  Equations (\ref{1}) $-$
(\ref{2}) above are the principal equations of the present model. The
total number of particles $N$ of the system is given by the number
equation \begin{eqnarray}\label{6} N=N_0+\int \frac
{\rho(E)dE}{\exp[(E-\mu)/k_BT]-1}, \end{eqnarray} where $N_0 \equiv \int
|\Psi(r)|^2 d^{\cal D}r$ is the total number of particles in the
condensate.
The critical temperature $T_0$ is obtained as the solution of Eq.
(\ref{6}) with $N_0$ and $\mu$ set equal to 0. The semiclassical density
of states $\rho(E)$ of noncondensed particles is given by \cite{7,9}
\begin{equation} \rho(E)=\frac{2\pi m^{{\cal D}/2}}{(2\pi\hbar)^{\cal
D}}\int_{V_{\mbox{eff}}(r)<E}[8(E-V_{\mbox{eff}}(r))]^{({\cal D}-2)/2}
d^{\cal D}r.  \label{7} \end{equation}

The average single-particle energy of the noncondensed particles
is given by \cite{7}
\begin{equation}\label{8} \langle E \rangle_{\mbox{nc}}=\int \frac
{E\rho(E)dE}{\exp[(E-\mu)/k_BT]-1}. \end{equation} 
The kinetic energy of the condensate is assumed to be
negligible and its interaction energy per particle is given by $
\langle E
\rangle _{\mbox{c}}=(g/2)\int \Psi^4(r) d^{\cal D}r$. The quantity of
experimental interest is the average  energy  $ \langle E
\rangle = [\langle E \rangle _{\mbox{nc}}(N-N_0)/2+\langle E \rangle
_{\mbox{c}}]/N, $ which we calculate in the following. The specific heat
is defined by $C=d\langle E \rangle/dT$ \cite{7}.

Equations (\ref{1}) $-$ (\ref{7}) are to be solved iteratively. The
iteration is started at a fixed temperature with a trial chemical
potential $\mu$ using $n_1(r) =0$. Then $\Psi(r)$ and $V_{\mbox{eff}} (r)$
are calculated using Eqs. (\ref{1}) and (\ref{2a}). With these results
$n_1(r), \Psi(r),$ and $V_{{\mbox{eff}}}(r)$ are recalculated. This
procedure is repeated until desired precision is obtained. The results for
the lowest order of iteration with $n_1(r)=0$ will be denoted by $I=1$,
and successive orders by $I=2,3,...$ etc. With the solutions $\Psi(r)$ and
$V_{{\mbox{eff}}}(r)$ so obtained, the density of states $\rho(E)$ of Eq. 
(\ref{7})  is calculated.  Then it is seen if they satisfy the number
equation (\ref{6}). If Eq. (\ref{6})  is satisfied the desired solution is
obtained. If not, the initial trial $\mu$ is varied until the number
equation is satisfied. In each order of iteration we calculate the
condensate fraction $N_0/N$ and  energy $\langle E \rangle$.

\section{Numerical Study}

First we consider the three-dimensional case. In this case the coupling
$g$ is given by $g=4\pi \hbar^2 a/m$, where $a$ is the scattering length. 
Usually, a dimensionless coupling is introduced via $\eta\equiv (mg/\pi
\hbar^2)/a_{\mbox{ho}} = 4 a/a_{\mbox{ho}}$, where
$a_{\mbox{ho}}=\sqrt{\hbar/m\omega }${}. The semiclassical model under
consideration should break down as either $\eta$ or $N$ is increased. This
will  violate the condition of weak interaction and diluteness. In
the case of experiment on BEC of $^{87}$Rb, $\eta = 0.025$ and
$N=40000$ \cite{1,7}. To test our calculational scheme, first we solve the
present model for $\eta =0.025$ and $N=40000$. Our results are very
similar to those of Ref.  \cite{7}. The small difference between
these two calculations is due to the use of an isotropic harmonic
oscillator potential in this work and an anisotropic potential in Ref. 
\cite{7}.  In this case specific heat is positive at all temperatures. 
Next we increase $\eta$ and $N$. We find that as $\eta$ and $N$ are
increased, the  energy develops a maximum at a temperature below
$T_0 $. Consequently, the specific heat becomes negative for an interval
of temperature above this maximum.  To show this violation in a pronounced
way here we show the results for the following three cases: (a) $\eta
=0.1$, $N=10^6$, (b)  $\eta =0.1$, $N=10^7$, and (c) $\eta=0.5, N=10^6$. 
In the case of $^{23}$Na, the experimental $N$ was as high as 10$^7$
\cite{2}. As the interaction is repulsive in both $^{87}$Rb and $^{23}$Na,
it should be possible to have 10$^7$ atoms in a BEC of $^{87}$Rb
under favorable experimental conditions. Hence these values of the
parameters are within the present  experimental scenario.

\vskip -3.8cm
\postscript{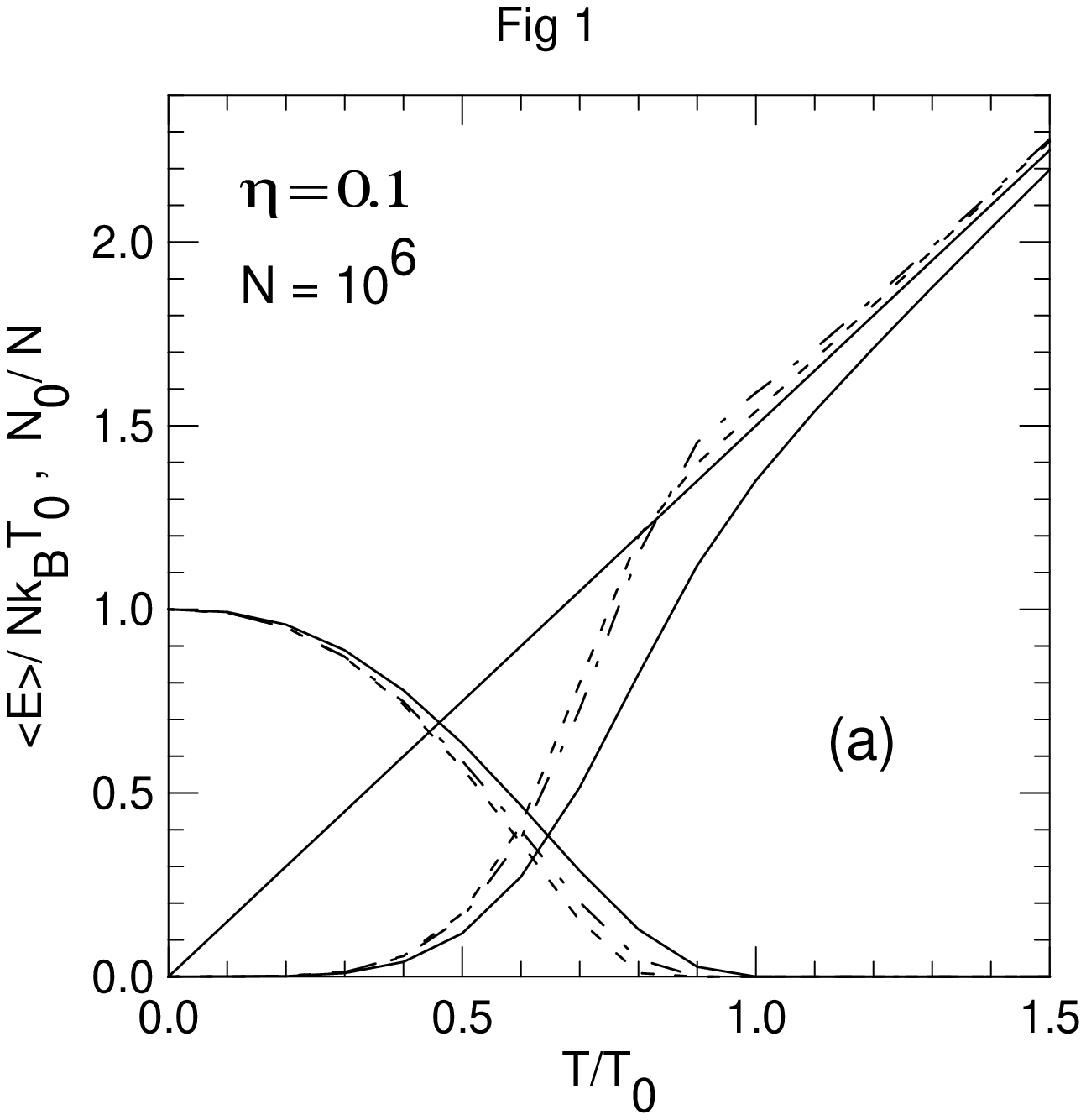}{1.0}    
\vskip -2.1cm

\vskip .4cm

\vskip -3.8cm
\postscript{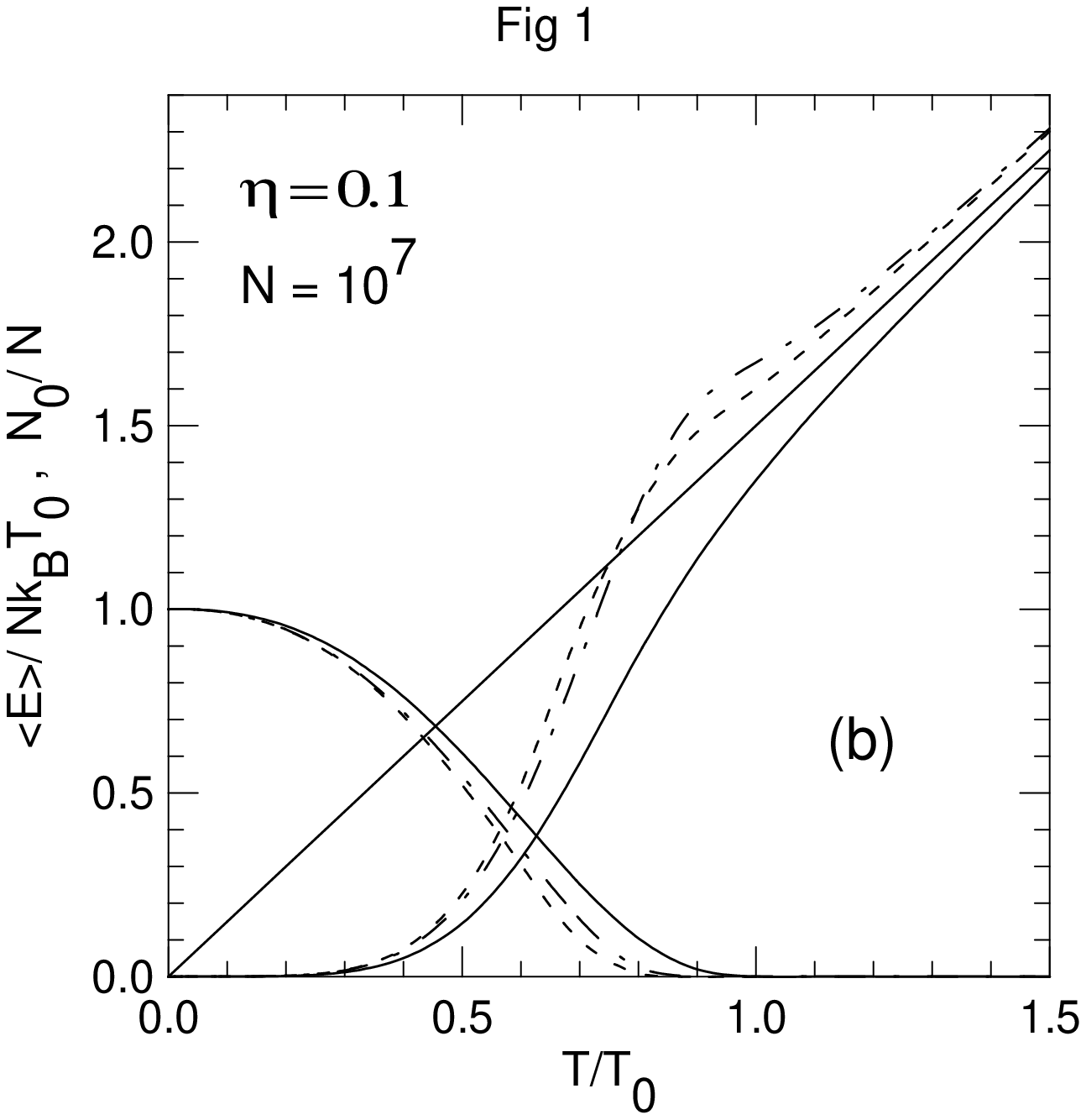}{1.0}    
\vskip -2.1cm

\vskip .4cm

\vskip -3.8cm
\postscript{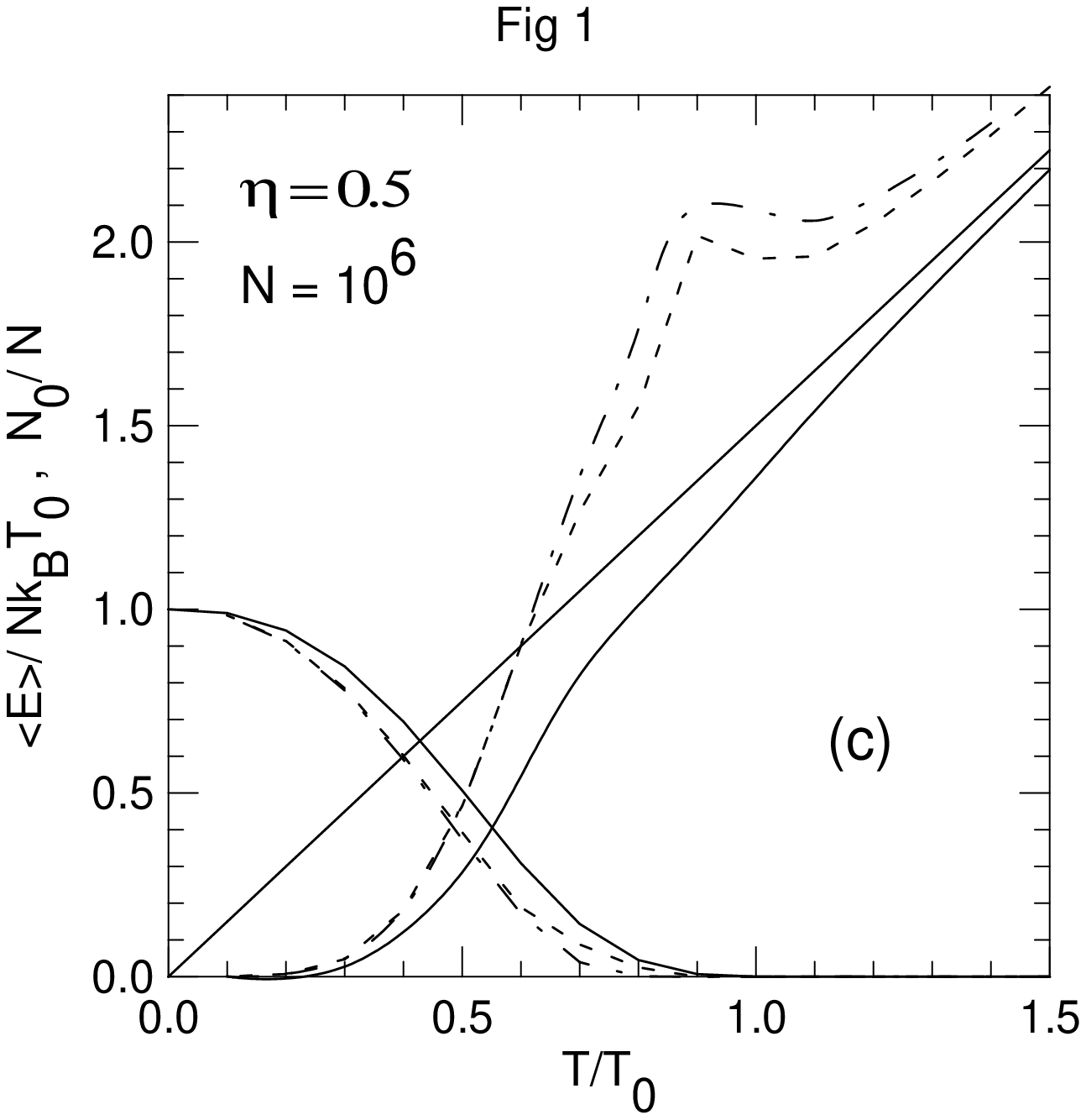}{1.0}    
\vskip -2.1cm

\vskip .4cm

{ {\bf Fig. 1.} Condensate fraction $N_0/N$ and  energy $\langle E \rangle
/Nk_BT_0$
in three dimensions as a function of $T/T_0$ for (a) $\eta =0.1$
and $N=10^6$,  (b) $\eta =0.1$ and $N=10^7$, and (c) $\eta =0.5$ and
$N=10^6$ for iterations
$I =1$ (full line), 2 (dashed-dotted line), and 4 (dotted line). The
straight line represents the classical Maxwell-Boltzman
result of energy. }

In Figs.
1 (a), (b), and (c) we plot $\langle E \rangle /Nk_BT_0$ and $N_0/N$
versus $T/T_0$ for different iterations for the above three cases.  We
find that the energies are acceptable under conditions of diluteness and
very weak interactions, but as $\eta$ and $N$ increases the average
 energy of the system may have a maximum leading to a negative
specific heat above the maximum.  From Figs. 1(a), (b), and (c) we find
that this violation happens in the case (c) which has the largest $\eta$
and $N$.  We verified that for sufficiently large values of $\eta$ and $N$
the lowest-order energy also leads to negative specific heat.

Next we consider the two-dimensional case. In this case, in analogy with
the three-dimensional case, a dimensionless coupling is introduced by
$\eta \equiv (mg/\pi\hbar^2)$. This coupling is already dimensionless,
whereas we needed to divide it by $a_{\mbox{ho}}$ in three dimensions to
make it dimensionless.  First we repeat the calculations reported in Ref. 
\cite{9} and our results are in agreement with that study. In addition, in
agreement with our finding in three dimensions, we find that for small
values of $\eta $ and $N$ the equations of the model converge well and
lead to acceptable values for condensate fraction $N_0/N$ and
energy $\langle E \rangle /Nk_BT_0$. For larger values of $\eta$ and/or
$N$, the condensate fraction $N_0/N$ is quite acceptable with a
temperature dependence similar to that in three dimensions.  However, the
 energy produces a maximum as $N$ and $\eta$ increase. Hence we
shall be limited to a consideration of  energy only. 

As there is no experimental guideline for probable values of $N$ and
$\eta$ in two dimensions, as in Ref. \cite{9}, we consider $N=10^5$ and
$\eta = 0.1,$ 1, and 10. In Fig. 2 we plot the temperature dependence of
average  energy $\langle E \rangle/Nk_BT_0$ for different
iterations.  The lowest-order results for $\eta =0.1$ and 1 are in
agreement with those of Ref. \cite{9}. The classical Maxwell-Boltzmann
result for a noninteracting gas is also shown in Fig. 2. For a weakly
interacting gas with $\eta = 0.1$, the energies for all orders of
iteration are acceptable. For a stronger interaction with $\eta =1$, the
lowest-order energy is acceptable. However, in this case the energies for
all orders of iteration produce maxima leading to negative specific heat. 
For $\eta = 10 $ the lowest-order result already leads to a negative
specific heat.

\vskip -3.8cm
\postscript{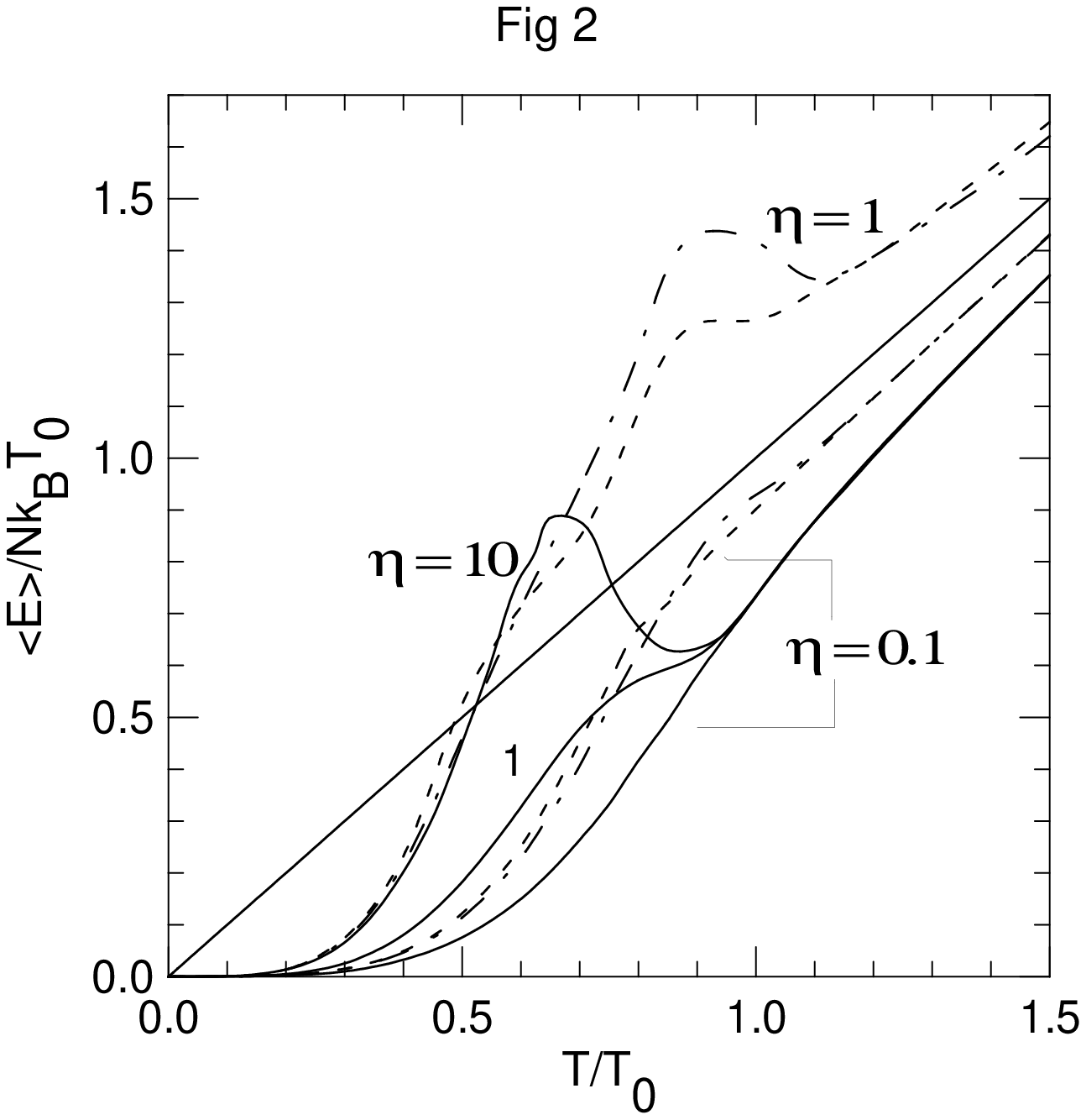}{1.0}    
\vskip -1.7cm

{ {\bf Fig. 2.}    Energy  $\langle E \rangle/Nk_BT_0$ in two
dimensions as a function of
$T/T_0$ for $N=10^5$ and  
$\eta =0.1$, 1, and 10 for iterations $I=1, 2$ and 4.  Notations are the
same as in Fig. 1. The curves are labeled by the values of $\eta$. }

In addition to the trouble discussed above, the mean-field model may
exhibit another unacceptable behavior. We see in Figs. 1 and 2, that the
 energy the system could be larger than the corresponding classical
Maxwell-Boltzmann result for a noninteracting ideal gas.  We recall that
for BEC to materialize, the energy of the condensate should be smaller
than the corresponding energy of the noncondensed system. However, one
should consider the correction to the classical result above due to the
presence of the trap and the interatomic interaction. In all our
calculations we have noted that $k_BT_0 >> \hbar \omega $, so that for
temperatures close to $T_0$ considered above the correction to the
classical energy due to the presence of the trap can be neglected. The
same is also true for the interatomic interaction at temperatures close to
$T_0$, for the values of $\eta$ considered in this study. Consequently,
the  energy of the condensate of the mean-field model could be
larger than the classical result signaling a breakdown of the model.

It is well known that the semiclassical model requires $
k_BT_0 >> \hbar \omega$ and the Thomas-Fermi approximation does not hold
for very small number  of atoms in the condensate. We have verified that
for the cases
studied here $k_BT_0$ is typically 50 to 100 times larger than $\hbar
\omega$ and Thomas-Fermi approximation apparently seems to be  valid for
the condensates of
size $N\sim 10^{6}$ to $10^7$ considered here. The diluteness condition
for the validity of the GP equation $na^3 << 1$ is also valid in three
dimensions, where $n$ is the density. For the case considered in Fig. 
1(a), $N=10^6$, $a=50$ nm, for a condensate of typical dimension 100
$\mu$m, $na^3 \sim 10 ^{-3} <<1$. The values of $na^3$ are larger for
Figs.
1(b) and 1(c). Nevertheless, we find that 
these values of diluteness and coupling
set a limit to the applicability of the semiclassical mean-field two-fluid
models for 
studying BEC thermodynamics.

\section{Conclusion}

In conclusion, we reexamined the problem of BEC under the
action of a trap potential using a two-fluid mean-field model \cite{7,9}
in both two and three dimensions. We employed an iterative solution scheme
of the system of equations. Although the system leads to rapid convergence
for a weakly interacting dilute system, with the increase of coupling and
particle number the iterative scheme leads to physically unacceptable
results for thermodynamic observables. Specifically, this may lead to a
maximum in  energy responsible for a negative specific heat.  In
three dimensions the breakdown of the mean-field model happens for values
of coupling and particle number, which are not so remote from present
experimental scenario for repulsive interatomic interaction.  In addition,
the energy of the condensate could be larger than the corresponding
classical energy of the system, which is another independent unacceptable
result of the model.  The larger values of the coupling $\eta $ and $N$
considered in this work possibly sets a limit to the applicability of the
mean-field equations for BEC.  Summarizing, the most important
finding of this study is that the iterative solution of the mean-field
model of Refs.  \cite{7,9} for a  condensate may lead to unphysical
thermodynamical properties for medium to large coupling and number of
particles. 

Despite  the above deficiency of the mean-field two-fluid
thermodynamical models, they continue to be very useful 
in many cases. The virtue of these models is the
simplicity and ability to yield results in agreement with experiment for
weak interactions and dilute systems. 
As a 
theoretically sound description of the BEC
thermodynamics seems to be unmanagably complicated, these simple
mean-field
two-fluid models remain as  attractive simple alternatives
to study 
thermodynamical properties 
provided that  proper attention is paid to remain inside the domain of
their validity.

AG thanks Prof. T. Frederico and Prof. L. Tomio for an introduction to the
subject of BEC.  The work is supported in part by the Conselho
Nacional de Desenvolvimento Cient\'\i fico e Tecnol\'ogico and Funda\c
c\~ao de Amparo \`a Pesquisa do Estado de S\~ao Paulo of Brazil.

\end{document}